# A state-space catch-at-length assessment model for redfish on the Eastern Grand Bank of Newfoundland reveals large uncertainties in data and stock dynamics

Noel G. Cadigan[*],
Centre for Fisheries Ecosystems Research, Fisheries and Marine Institute of Memorial University,
St. John's, NL, Canada, A1C 5R3
Noel.cadigan@mi.mun.ca

Andrea M. Perreault,
Science Branch, Fisheries and Oceans Canada,
St. John's, NL, Canada, A1C 5X1
andrea.perreault@dfo-mpo.gc.ca

Hoang Nguyen,
Centre for Fisheries Ecosystems Research, Fisheries and Marine Institute of Memorial University,
St. John's, NL, Canada, A1C 5R3
Hoang.Nguyenthe@mi.mun.ca

Jiaying Chen,
Centre for Fisheries Ecosystems Research, Fisheries and Marine Institute of Memorial University,
St. John's, NL, Canada, A1C 5R3
Jiaying.Chen@mi.mun.ca

Andres Beita-Jimenez,
Centre for Fisheries Ecosystems Research, Fisheries and Marine Institute of Memorial University,
St. John's, NL, Canada, A1C 5R3
Andres.Beita-Jimenez@mi.mun.ca

Natalie Fuller
Centre for Fisheries Ecosystems Research, Fisheries and Marine Institute of Memorial University,
St. John's, NL, Canada, A1C 5R3
Natalie.Fuller@mi.mun.ca

Krista Ransier
Centre for Fisheries Ecosystems Research, Fisheries and Marine Institute of Memorial University,
St. John's, NL, Canada, A1C 5R3
Krista Ransier@mi.mun.ca

[*]Corresponding Author

**Abstract**

We developed a state-space age-structured catch-at-length (ACL) assessment model for redfish in NAFO Divisions 3LN. The model was developed to address limitations in the surplus production model that was previously used to assess this stock. The ACL model included temporal variations in recruitment, growth, and mortality rates, which were limitations identified for the surplus production model. Our ACL model revealed some important discrepancies in survey and fishery length compositions. Our model also required large population dynamics process errors to achieve good fits to survey indices and catch estimates, which also demonstrated that additional understanding of these data and other model assumptions is required. As such, we do not propose the ACL model to provide management advice for 3LN redfish, but we do provide research recommendations that should provide a better basis to model the 3LN redfish stock dynamics. Recommendations include implementing sampling programs to determine redfish species/ecotypes in commercial and research survey catches and improving biological sampling for maturity and age.

**Introduction**

Redfish (*Sebastes spp.,* Sebastidae) on the Eastern Grand Bank of Newfoundland (NAFO Divisions 3LN) historically supported an important commercial fishery, with an average landed catch of approximately 26 000 tonnes from 1959 to 1993. Landings declined substantially in 1994, and a moratorium was declared in 1998. In 2011, commercial fishing resumed with total landings of 3 672 tonnes, and landings were estimated at 10 172 tonnes in 2021 (Rogers et al., 2022). Redfish are long-lived, slow-growing, late-maturing, and tend to produce large year classes episodically (Cadigan et al., 2022). In addition, there are two distinct species of redfish in 3LN, *S. mentella* and *S. fasciatus,* which are morphometrically very similar (Ávila de Melo et al., 2020); however, species are not distinguished in the survey and commercial fishery catches. As such, standard assessment approaches that have been successful for other stocks have often performed poorly for 3LN redfish due to the complicated characteristics of this stock complex.

The assessment model used from 2008 to 2020 for 3LN redfish (Ávila de Melo et al., 2020) was a logistic surplus production model (SPM) implemented using the ASPIC software package (Prager, 1994, 2016). The model was based on landings estimates from 1959 onward and nine indices of stock size. Since 2014, maximum sustainable yield (MSY) in the model was fixed at the 1960-1985 average catch of 21 000 tonnes. Although it is unusual to fix MSY when estimating stock assessment model parameters, this constraint was presumably used to reduce confounding among SPM parameter estimates and force the MSY value to be consistent with historic landings. However, even with the fixed MSY, the SPM was rejected at the 2022 stock assessment due to strong patterns in the model residuals (Rogers et al., 2022). Lack of model fit was suggested to be primarily driven by sporadic recruitment and changes in length compositions of the stock over time, which have been previously documented for 3LN redfish (Rogers et al., 2022, Perreault et al., 2022).

SPMs are known to perform poorly when there are time-varying changes in age and size dynamics in growth, fishery selectivity, recruitment, and natural mortality rates (Punt and Szuwalski, 2012), and a variety of extensions have been proposed to address these issues. Including process errors in the SPM population dynamics equation has been shown to partially account for time variations in the production function (e.g., Rankin and Lemos, 2015, Pedersen and Berg, 2017, Winker et al., 2018). Including a time-varying intrinsic growth rate performed better when fishing mortality and predation mortality changed over time (Nesslage and Wilberg, 2019). Distinguishing between exploitable biomass and spawning biomass in a Bayesian framework outperformed traditional SPMs (Winker et al., 2020). However, the sensitivity of SPMs to episodic recruitment has not been well studied and remains an unresolved source of uncertainty in the application of SPMs to 3LN redfish.

To address some of these shortcomings, we develop an integrated state-space age-structured catch-at-length (ACL) model for 3LN redfish that utilizes the available information on fishery total catch-at-length and abundance indices-at-length from multiple research vessel surveys. Including length composition data from surveys and fishery sampling provides important information about recruitment and mortality rates over time. SPMs do not utilize size composition information and these models must infer productivity dynamics based only on how total catches have affected survey indices of exploitable stock biomass. The ACL model estimates fishing mortality rates at age, recruitment deviations, and separate length-based catchability parameters for each survey. The ACL model also includes process errors which can partially account for changes in natural mortality rates. For these reasons, the ACL model is a better model of redfish stock dynamics and its uncertainties compared to SPMs. The ACL is a simple extension compared to more complicated age-length structured models (e.g., Zhang and Cadigan, 2022) or length-only models.

**Materials and Methods**

A state-space stock assessment model consists of two parts: a process model and an observation model; the process model describes the underlying population dynamics of the stock, and the observation model describes how catch and survey data depend on the unobserved population dynamics (e.g., Aeberhard et al., 2018).

**The process model:**

Our ACL model was developed from the standard age-structured model for fish population dynamics (e.g., Quinn and Deriso, 1999),

$$N_{a+1,y+1} = N_{a,y} \times e^{-Z_{a,y}} \times e^{\delta_{N_{a,y}}}, \quad (1)$$

where $N_{a,y}$ are the numbers of fish in each age class, $a$, in year, $y$. The process error terms are independent $\delta_{N_{a,y}} \sim N(0, \sigma_{pe})$. The total mortality rate is

$$Z_{a,y} = F_{a,y} + M_{a,y}, \quad (2)$$

where $F_{a,y}$ and $M_{a,y}$ are the fishing mortality and natural mortality rates, respectively. We assumed that redfish at ages 1-2 years old are not targeted by the fishing gear (i.e., $F_{a=1,y} = F_{a=2,y} = 0$ for all $y$). We modeled $F$ as the product of the mean F ($\mu_{F,y}$) of ages 3 onwards and a deviation $\delta_{F_{ay}}$,

$$log(F_{a,y}) = \log(\mu_{F,y}) + \delta_{F_{a,y}}, \quad \mu_{F,y} = \begin{cases} \mu_{F,mort}, & y = 1998,\dots,2009, \\ \mu_{F,fish}, & \text{other years}, \end{cases} \quad (3)$$

where $\delta_{F_{a,y}}$ was modeled as a correlated AR(1) process, separable across ages and years, with parameters $\sigma_F^2$, $\varphi_{F_a}$ and $\varphi_{F_y}$ to estimate. We assumed $\mu_{F,y}$ was the same for the years when there was a fishing moratorium (i.e., $\mu_{F,mort}$), but a different value ($\mu_{F,fish}$) for the other years when there was a TAC>0. We accounted for an abrupt change in $F$ due to the fishing moratorium by using a different mean $F$ parameter for that period. Note that although the fishing moratorium was in place during 1998-2010, we did not include 2010 in the mean $F$ parameter block for the moratorium because the landings in 2010 (4 120 tonnes) were substantially larger than other years and were more like those in 2011-2015 (see Rogers et al., 2022).

Natural mortality rates ($M$s) were predicted as a function of body weight using the Lorenzen method (Lorenzen, 1996),

$$M_a = M_o \times W_a^{-0.305}, \quad (4)$$

where $W_a$ is the stock weight-at-age, $M_o$ is a scaling parameter and -0.305 is the value for ocean systems from Lorenzen (1996), similar to Miller and Hyun (2018) and Kumar et al. (2020). $M$ was assumed to be time invariant, so $M_{a,y} = M_a$ for all years. We solved for $M_o$ by re-scaling $M_a$ using $M_{median}\left(\frac{M_a}{Min(M_a)}\right)$ where $M_{median} = 0.125$ is the median $M$ derived from Cadigan et al. (2022). Weight-at-age was derived using stock length-weight and length-age relationships,

$$W_a = \alpha L_a^\beta, \quad (5)$$

where $L_a$ are internally estimated using the Von Bertalanffy growth model (Equations 7 & 8 below). The length-weight parameters $\alpha$ and $\beta$ were derived from length-weight estimates from the EU-Spain research vessel surveys (see Data Description for details). Cadigan and Campana (2017) found that increases in the length of 3LN redfish were negligible after reaching age 10; therefore, our model runs from ages 1-10+, where 10+ represents the plus group. Model sensitivity runs for ages 15+ indicated little differences in model results.

The recruitment is the abundance of fish in age class 1 in each year and was modeled as the product of mean recruitment $\mu_{R_y}$ and annual multiplicative deviations $\delta_{R_y}$,

$$N_{1,y} = \mu_{R_y} \times exp(\delta_{R_y}), \quad (6)$$

where $\delta_{R_y} \sim AR(1)$ with parameters $\sigma_R^2$ and $\varphi_R$ to estimate.

*Probability of length-at-age*

The model is age-structured; however, age information for 3LN redfish is not readily available but catch-at-length and survey-index-at-length data are. Therefore, we used a stochastic growth model to convert model numbers-at-age to numbers-at-length, which we could then fit to. We started from the original form of the Von Bertalanffy growth model (Von Bertalanffy, 1938),

$$L_a = L_\infty - (L_\infty - L_0)e^{-Ka}, \quad (7)$$

where $L_a$ is the expected mean length-at-age, $L_\infty$ is the asymptote for the model of average length-at-age, $K$ is the growth rate parameter, and $L_0$ is the mean length at age zero (i.e., at birth). $L_0$ is a difficult parameter to estimate so we fixed $L_0$ at 0.58 cm, based on Penney & Evans (1985).

The length information was measured for 1 cm length bins, and we assume that the reported length is a mid-point of the length bins. Hence, a fish at length bin $l$ will have a length $L \in (l - 0.5$ cm, $l + 0.5$ cm). We used the cumulative distribution function (CDF) $P(a < X \leq b) = F_X(b) - F_X(a)$ to compute the probability that a fish is in length bin $l$, given its age. We assumed that length-at-age is normally distributed within each length bin with mean $L_a$ and that the standard deviation is proportional to the mean, denoted $\tau L_a$, i.e., $L_a \sim N(L_a, \tau L_a)$. Using the CDF of the standard normal distribution $Z \sim N(0,1)$, $\Phi(z)$, the probability that a fish is in length bin $l$ given the fish is age $a$ at the beginning of the year is

$$P(L_a \in l) = \Phi\left(\frac{l - L_a + 0.5}{\tau L_a}\right) - \Phi\left(\frac{l - L_a - 0.5}{\tau L_a}\right). \quad (8)$$

In the model, $P(L_a \in l)$ is denoted as $P_{l,a}$, and applying $P_{l,a}$ to equation (1), we calculated the number of fish for each length bin in each year ($N_{l,y}$) as follows,

$$N_{l,y} = \sum_a N_{a,y} \times P_{l,a}. \quad (9)$$

**The observation model:**

We used the Baranov catch equation for the relationship between the population catch $C_{a,y}$ and stock size $N_{a,y}$,

$$C_{a,y} = N_{a,y} \times (1 - e^{-Z_{a,y}}) \times \frac{F_{a,y}}{Z_{a,y}}. \quad (11)$$

Assuming the distribution of length-at-age in the catch data is the same as that of the stock, we computed the number of fish caught for each length bin in each year ($C_{l,y}$). We assumed the catch occurs at mid-year. We also assumed constant growth rates throughout the year and estimated the distribution of length-at-age at mid-year using $P_{l,a}$ but modified by adding 0.5 to the ages in Equation (9), which we denoted as $P_{c,l,a}$. The model catch-at-length was

$$C_{l,y} = \sum_a C_{a,y} \times P_{c,l,a}. \quad (12)$$

We added a lognormal observational error term, $\varepsilon_{C,l,y}$, to equation (12) to describe the relationship between the predicted and observed catch ($C_{o,l,y}$),

$$C_{o,l,y} = C_{l,y} \times e^{\varepsilon_{C,l,y}}. \quad (13)$$

The $\varepsilon_{C,l,y}$ sampling errors were multivariate normal (MVN) with mean zero and lag-1 autoregressive correlation $\rho_C$ among length bins each year, but otherwise independent between years; that is, $corr(\varepsilon_{C,l,y}, \varepsilon_{C,k,y}) = \rho_C^{|l-k|}$ and $Var(\varepsilon_{C,l,y}|\varepsilon_{C,l-1,y}) = (1 - \rho_C^2)\sigma_C^2$.

There are three surveys in 3LN that occur at various times of the year. Denoting $N_{s,a,y}$ as the number of fish in the stock at the time of survey $s$, the relationship between stock size at the beginning of the year ($N_{a,y}$) and at the time of the survey ($N_{s,a,y}$) is given by,

$$N_{s,a,y} = N_{a,y} \times exp\left(-f_s \times Z_{a,y}\right), \quad (14)$$

where $f_s$ represents the fraction of the year the survey occurs in (e.g., $f_s = 8/12$ for a survey in August). The number of fish for each length bin in each year was estimated by

$$N_{s,l,y} = \sum_a N_{s,a,y} \times P_{s,l,a}, \quad (15)$$

where $P_{s,l,a}$ is the probability at length at the time of a survey ($P_{s,l,a} = P_{l,a+f_s}$) and $P_{l,a} = P(L_a \in l)$ defined in Eqn. (8) The expected survey index, $E(I_{s,l,y})$, is the product of length-based catchability, $q_{s,l}$, and the number of fish available to the survey,

$$E(I_{s,l,y}) = q_{s,l} \times N_{s,l,y}. \quad (16)$$

Catchability was modeled as a lognormal random walk,

$$q_{s,l+1} = q_{s,l} \times e^{\varepsilon_{q,l}}, \quad (17)$$

with $\varepsilon_{q,l} \sim N(0, \sigma_{q_{s,l}})$. To avoid wildly fluctuating estimates at the largest and smallest sizes, $\sigma_{q_{s,l}}$ was fixed at 0.20 for lengths less than 16cm and greater than 31cm. Estimates of $q_{s,l}$ for lengths greater than 45cm were fixed at the estimate of 45cm. To avoid confounding between $q_{s,l}$ and $N_{s,l,y}$, we added a medium prior on the DFO Fall research vessel (RV) survey $q_{s,l}$, where we assumed that $q_{s,l}$ was approximately 1 for lengths of 20, 22 and 24. We used an observational error term, $\varepsilon_{s,l,y}$, to model sampling variability in the survey observations,

$$I_{s,l,y} = E(I_{s,l,y}) \times e^{\varepsilon_{s,l,y}}. \quad (18)$$

The $\varepsilon_{s,l,y}$ sampling errors were multivariate normal (MVN) with mean zero and lag-1 autoregressive correlation $\rho_I$ among length classes each year, and conditional standard deviations $\sigma_{s,I}$ that were estimated separately for each survey *s*. For simplicity, the autocorrelation was assumed to be the same for all surveys.

Additional model outputs of interest are:

$$B_{l,y} = N_{l,y} \times SW_l \text{ and } B_y = \sum_l B_{l,y} \quad (19)$$

$$SSB_{l,y} = B_{l,y} \times Mat_l \text{ and } SSB_y = \sum_l SSB_{l,y} \quad (20)$$

$$C_y = \sum_l C_{l,y} \times CW_l \quad (21)$$

$$H_y = \frac{C_y}{B_y} \quad (22)$$

where $B_{l,y}$ is the stock biomass in year *y*, $SW_l$ and $CW_l$ are the mean stock and catch weight-at-length, respectively, $Mat_l$ is the female maturity-at-length (detailed below), $C_y$ is the catch in weight computed for year *y*, and $H_y$ is the harvest rate.

**Data description**

Commercial catch-at-length data are available for 3LN redfish for the years 1990-2021 (see Table 1 for additional details). Catches greater than 40cm occurred infrequently, so we aggregated data greater than 40cm into a ≥40cm bin. Similarly, catches less than 20cm were aggregated in a ≤20cm bin. Survey indices were obtained from the 3LN Canadian and 3LN EU-Spain surveys. The converted Canadian survey data (see Perreault et al. 2022 for details on comparative fishing) are available for 1991-2020 for the fall and 1992-2019 for the spring surveys. The replacement of larger mesh-sized nets in the summer of 1995 resulted in indices of small-sized fish since then that are not comparable with indices prior to the replacement; therefore, we excluded fish less than 20cm in length during 1991-1994 for the fall survey and 1992-1995 for the spring survey. As in the commercial data, redfish greater than 40cm occurred infrequently, so we aggregated indices greater than 40cm in a ≥40cm length bin. Survey indices at lengths less than 10cm were aggregated in a ≤10cm length bin for fall and spring surveys since 1995.

Converted EU-Spain survey data were available for 3LN for the years 2004-2019 and 1997-2019, respectively; however, we only used survey indices from 3N for the model estimation. The EU-Spain survey is conducted outside of the 200nm zone and has only covered approximately 15% of the total 3L redfish area over time (Supplementary Figure 1). As such, these data were not expected to provide a reliable indicator of stock trends and were not considered further in the model. Abundance greater than 40cm occurred infrequently, so we aggregated data greater than 40cm in a ≥40cm length bin, and abundance less than 10cm were aggregated in a ≤10cm length bin.

Length-weight parameters (i.e., $W(l) = \alpha l^\beta$) were available from the 3LN Portuguese commercial sampling data during 1990-2019 (see Ávila de Melo et al., 2020) and the 3LN EU-Spain RV data during 1997-2021. For each data set, estimates of mean weight-at-length for each year were derived from yearly $\alpha$ and $\beta$ parameters, and a length-weight model was fit to the predictions of mean weight-at-length for all years. Due to data limitations, this was considered a reasonable approach to estimate a single alpha and beta parameter for stock ($\alpha_s = 0.009, \beta_s = 3.11$) and catch ($\alpha_c = 0.042, \beta_c = 2.69$) weights. Mean catch weights (CW) and stock weights (SW) at length (Equations 19 and 21) were estimated using these alpha and beta parameters, and were assumed to be time invariant (i.e., the same mean stock weight at length across all years).

**Parameter estimation**

The key to solving a mixed-effect model is to compute the joint likelihood function of observed and unobserved states,

$$L_{joint}(\theta, \mathcal{D}, \Psi) = p_\theta(\mathcal{D}|\Psi)p_\theta(\Psi), \quad (23)$$

where $\mathcal{D}$ represents the observed data (i.e., commercial catch, survey abundance indices-at-length), $\Psi$ is the vector of random effects, and $\theta$ is the vector of fixed-effect parameters. A good approach to estimate $\theta$ is based on maximizing the marginal likelihood,

$$L_{marginal}(\theta, \mathcal{D}) = \int p_\theta(\mathcal{D}|\Psi)p_\theta(\Psi)d\Psi. \quad (24)$$

The high-dimensional integral in Eqn. (24) can be solved via the Laplace approximation (see e.g., Aeberhard et al., 2018 for details). We used the TMB (Kristensen et al., 2016) and nlminb packages in the R environment (R-Core-Team, 2021) to minimize the negative marginal loglikelihood. The model parameters, including fixed and random effects, are described in Table 2.

**Model exploration**

We describe below that we were not successful in developing a robust and reliable assessment model using the assessment inputs and assumptions described above. There are serious discrepancies in the data and our assumptions. Therefore, in this paper we do not present a thorough model sensitivity and selection process; however, to better understand potential drivers of possible model misspecification, we considered the following:

1) Fitting the model for various survey catchability formulations. The model formulation described above estimated catchability for each survey and for lengths up to 45cm. We considered various formulations that essentially force flat-topped catchability at larger sizes: base case, fixing max estimate at 45cm; q40, fixing max estimate at 40cm; q30, fixing max estimate at 30cm; and q20, fixing max estimate at 20cm. We use the Akaike information criterion (AIC) in model selection (note we do not use BIC since the number of fixed effect parameters do not change in this model selection process).

2) Residual analyses. We examined two types of survey and catch residuals: Standardized (i.e., Pearson) residuals, which are (O-E)/SD, where O is the observed log index or log catch, E is the model prediction, and SD is the standard deviation of the prediction. The raw residuals O-E, which are correlated in our observation models, were also standardized using the Choleski factorization of their estimated covariance matrix (e.g., Houseman et al., 2004). We refer to these as Z-residuals. They are uncorrelated, whereas the standardized residuals are not. However, sometimes these Z-residuals can mask lack of fit to survey indices, which is why we prefer to examine both types of residuals.

3) Comparing our model output to the last published surplus production ASPIC model from 2019 (Ávila de Melo et al., 2020).

**Results**

In preliminary models we found that estimates of the Von Bertalanffy $K$ and $L_\infty$ parameters were highly confounded (i.e., negatively correlated) and poorly determined. Hence, in our initial ALK assessment model formulations we fixed $L_\infty = 52$ cm. Other estimated model parameters are listed in Table 3. The model predicted biomass was lowest during 1994-1999 but increased to a higher level after the mid-2000's (Figure 1, top left). Fishing mortality rates were highest in 1990-1993 (Figure 2, top left). Commercial catches were predicted to be targeting younger and smaller fish during 2010-2015 (Figure 2, bottom panels). Overall, the base model fit the commercial (Supplementary Figures 2 and 3) and survey (Supplementary Figures 4 and 5) length composition data well, except for a lack of fit at smaller sizes for both the commercial and survey data. Model predicted commercial length compositions follow observed trends well (Supplementary Figure 6). However, in some years there are survey year-effects (Supplementary Figure 5).

There are two indications of serious model misspecification. The first is the catchability at length estimates for the three surveys (Canadian fall and spring and EU-Spain 3N). When freely estimated (base model), the catchabilities are all strongly dome-shaped, peaking at approximately 20cm for all surveys (Figure *3*, solid line). This is unexpected for groundfish surveys and indicates that the surveys are not seeing larger redfish that are in the catch. This is also obvious by comparing the catch length compositions (Supplementary Figure 6) with the surveys (Supplementary Figures 7-9). The proportion of fish greater than 30cm in the total catch in each year (Supplementary Figure 10) indicate that in the early years of the model the commercial catches were often dominated by large fish (~40-60% of the total catch) in comparison to the surveys, where large fish comprised on average 10% of the total catch. Although the proportion of fish greater than 30cm in the total catch has decreased since the reopening of the commercial fishery in 2011 (~20% of the total catch), they are still less frequently caught in the survey catches.

The other indication of substantial model misspecification is the large value of the process error standard deviation, $\sigma_{pe}$ (Table 3). This value is usually less than 0.3 in state-space stock assessment models. The large value was required to fit the survey indices and catch statistics. Smaller values of $\sigma_{pe}$ resulted in a much worse fit to the data, which indicates that the reported catches and our assumptions about M have not described redfish population dynamics well, as indicated by the survey indices. Trends in process errors (e.g., in the year 2002, process errors are negative at most ages, and errors at ages 7-9 are mostly negative; Supplementary Materials Figures 11-12) also suggest that there are underlying factors not captured in the current model formulation. The values of $\sigma_I$ (Table 3) are high, which indicates that even with a large amount of process error, there is substantial lack of fit to the survey indices.

The best fitting model for the various survey catchability formulations (base case: fixing max estimate at 45cm; q35: fixing max estimate at 35cm; q25: fixing max estimate at 25cm, and q20: fixing max estimate at 20cm), was the q20 model, although the differences between the base, q25 and q35 models were negligible (Table 3). Overall, predicted recruitment was similar for all model formulations (Figure *5*). Catchability estimates were smaller as max *q* decreased (Figure 3), which resulted in lower biomass and larger fishing mortality estimates.

The 2019 ASPIC model (Figure *6*) predicted much lower biomass and larger harvest rates than the ACL model, although general trends in harvest rates were similar. Trends in biomass differed somewhat, although this is not entirely unexpected, since the 2019 ASPIC model fixed the MSY estimate at 21 000 tonnes. Trends from ASPIC model runs with MSY freely estimated more closely resembled the trends from the ACL model (A. Perreault personal communications, 2023).

Model convergence was an issue in sensitivity (i.e., profile likelihoods) and retrospective runs.

**Discussion**

The NAFO assessment model for 3LN redfish (i.e., a surplus production model, SPM) was rejected at the 2022 stock assessment due to strong patterns in the model residuals (Rogers et al., 2022). SPM's are known to perform poorly when there are time-varying changes in age and size dynamics in growth, fishery selectivity, recruitment, and natural mortality rates (M's). Episodic recruitment is common with redfish, which is an important deficiency of using an SPM for redfish management advice. We developed a novel age-structured catch-at-length model (ACL) that included more information about stock dynamics (i.e., temporal variation in recruitment and mortality rates) from fishery and survey length composition samples. The ACL model accounted for time-variation in fishery selectivity and recruitment, which the SPM did not. The ACL also included process errors which can partially account for changes in M's, although there is ambiguity about the efficacy of this (Aldrin et al., 2020). The ACL model fits revealed some important discrepancies in survey and fishery length compositions. Our ACL also demonstrated that our model assumptions did not provide good fits to survey indices and catch estimates, and large population dynamics process errors were required. As such, we do not propose the ACL model to provide management advice for 3LN redfish, but we do provide research recommendations to better model the 3LN redfish stock dynamics.

A main driver of the current lack of fit to the data is the mismatch between the length compositions of the survey indices and commercial catches. Heavily dome-shaped catchability estimates and catch and survey length compositions plots provide evidence that the commercial fishery is targeting larger fish than those seen in the research vessel surveys, more notably in the early 1990s. There is a potential that the commercial fishery may be targeting pockets of the *S. mentella* stock in 3LN that are the southern extensions of the 2+3K *S. mentella* stock, whereas the surveys in 3LN may be finding *S. fasciatus* more frequently. We suggest that S. *fasciatus* is the more common and widely distributed redfish species in 3LN. However, the 3LN commercial length frequencies seem more consistent with the sizes of redfish caught in the DFO fall surveys in 2+3K (unpublished results). If this is the case, then it should not be too surprising that the commercial catches of *S. mentella* in 3LN are not helping to explain the dynamics of S. *fasciatus* as indicated by the surveys, which is one explanation for why our ACL model produced such large process errors. More detailed spatial assessment models will be required to address this issue.

Recent research suggests an even more complex population structure, including three ecotypes within *S. mentella* and five populations of *S. fasciatus* across the Northwest Atlantic (Benestan et al., 2021). Ignoring stock and sub-stock structures has been shown to produce unsustainable or suboptimal harvesting advice and can potentially lead to misperception of the magnitude of fish productivity (e.g., Cadrin and Secor, 2009, Kerr et al., 2017). However, the current data available for 3LN redfish does not differentiate between species or subtypes, and species identification with the commercial and survey length composition data is difficult. Additional information to identify redfish species (i.e., genetic samples, detailed morphometrics) is required. Otherwise, a mixed-species assessment model such as ICES (2023) based on Albertsen et al. (2018) will be required to account for the population dynamics of the two redfish species, but we expect such a model will be difficult to fit and have high uncertainty without species composition data.

Redfish males and females exhibit differences in their growth rates and sizes at the same age (Saborido-Rey et al., 2004; Cadigan & Campana, 2017). Females generally grow faster and are larger than males of the same age. Additionally, female redfish have a longer lifespan than males, and they typically reach larger sizes at older ages (Stransky et al., 2005). However, our ACL model is sex-aggregated. These differences in growth rates and sizes between male and female redfish are important to consider when assessing stock status. A sex-specific model may give better estimates of stock dynamic if reliable sex-composition information is available.

When reliable age data are unavailable, length-structured models have often been used in place of simpler surplus production models to estimate productivity parameters of interest (e.g., fishing mortality rates and biomass; Punt et al., 2013) and this remains the main motivation for using our ACL model. It is important to note that we assumed that the distribution of size-at-age in the population is the same as the catch, which might not be true if mortality is length-dependent. In this case, a three-dimensional age-and-length structured catch-at-length model may be more appropriate (e.g., ALSCL; Zhang & Cadigan, 2022) and a potential interesting direction for future work.

Our ACL model began in 1990 and aggregated ages 10+. Redfish are a slow-growing, long-lived species, and a model that begins in 1990 may not adequately capture trends in population dynamics and response to harvest rates. Additionally, the estimate of mean length for the oldest ages has a large influence on the estimates of fishing mortality rates and biomass in integrated models, and as such incorrectly specifying the age 10+ group may have unintended impacts on the perceived stock status (Maunder & Piner, 2015). The ACL model sensitivity runs for ages 15+ provided little difference in model output, although this may indicate issues with the model formulation (i.e., heavily domed $q$s) and not robustness. We suggest that future work aim to expand catch-at-length data prior to 1990, to fit a model that can better capture the history of the stock, and further explore growth and age structure formulations. There are other historical indices of abundance that were used to fit SPM for 3LN redfish (Ávila de Melo et al., 2020) could also be included in a model that covers a longer timeframe.

The available data are an important limiting factor to our understanding of the 3LN redfish stock dynamics. Cadigan et al. (2022) provided a thorough review of the available information for Northwest Atlantic redfish, with concluding recommendations for future research. We highlight three recommendations from Cadigan et al. that we consider to be of high importance for 3LN redfish in order to improve our future modeling approaches for the stock: 1) develop and implement sampling programs to determine redfish species/ecotypes in both the commercial and survey sampling, 2) improve biological sampling to get a better understanding of maturity and age of the stock, and 3) investigate harvest strategies that are appropriate for the life history and variable population dynamics of redfish species/ecotypes in 3LN and the wider Northwest Atlantic. The stock of 3LN redfish is currently without a stock assessment model, and new research avenues should be of high priority to further the best available science to inform the management of the stock.

The model developed in this paper has provided interesting new insights into important assessment data (i.e., size compositions) previously not used in the assessment model for 3LN redfish. Initial model fits and data exploration provided evidence of differences in length compositions for the commercial and survey data. We suggest that future model formulations allow for a divisional and/or depth-based model. Beyond the scope of this model, we also suggest revisiting the spatial extent of the management units. Although state-space models allow for some immigration and emigration in and out of the stock area, ignoring substructures and spatial structures can, in some cases, lead to overfishing or suboptimal harvesting advice (e.g., Cadrin, 2020). Although our model is not yet ready for use in an assessment framework, we suggest that progress on our research recommendations should provide a good basis to develop a reliable assessment model for 3LN redfish.

**Acknowledgements**

Research funding was provided by the Ocean Frontier Institute, through an award from the Canada First Research Excellence Fund. Research funding to NC was also provided by the Ocean Choice International Industry Research Chair program at the Marine Institute of Memorial University of Newfoundland.

**Tables**

Table 1. Description of catch and swept-area abundance indices-at-length. All data are in one cm length bins.

| Source | Season | Duration | Sampling area |
|---|---|---|---|
| Catch | All | 1990 – 2021 | 3LN |
| Canadian surveys | Fall | 1991 – 2020 | 3LN |
| | Spring | 1992 – 2019 | 3LN |
| Spanish survey | Spring | 1997 – 2019 | 3N |

Table 2. Description of model parameters. SD – standard deviation

| | |
|---|---|
| **Acronyms** | |
| NAFO | Northwest Atlantic Fisheries Organization |
| SPM | Surplus production model |
| MSY | Maximum sustainable yield |
| ACL | Age-structured catch-at-length |
| AR(1) | Autoregressive order one |
| CDF | Cumulative distribution function |
| MVN | Multivariate normal |
| RV | Research vessel |
| AIC | Akaike information criterion |
| BIC | Bayesian information criterion |
| ALK | Age structured catch-at-length model |
| SPLY | Standardized proportion at length and year |
| ALSCL | Age-and-length structured catch-at-length model |
| SPSH | Spanish |
| **Index** | |
| $a$ | Age class |
| $y$ | Year |
| $l$ | Length class |
| $s$ | Survey |
| **Data** | |
| $f_s$ | Fraction of the year that the survey occurs |
| $\alpha$ | Weight-length condition coefficient |
| $\beta$ | Weight-length allometric coefficient |
| **Process model** | |
| $N_{ay}$ | Numbers of fish in each age class $a$ in year $y$ |
| $Z_{ay}$ | Total mortality rate at age class $a$ in year $y$ |
| $F_{ay}$ | Fishing mortality rate at age class $a$ in year $y$ |
| $M_{ay}$ | Natural mortality rate at age class $a$ in year $y$ |
| $M_o$ | Scaling parameter |
| $W_a$ | Stock weight-at-age |
| $P_{la}$ | Probability that a fish is in length bin $l$ given the fish is age $a$ |
| *Fixed Effects* | |
| $\mu_{R_y}$ | Mean recruitment |
| $\sigma_R$ | SD of recruitment deviations |
| $\varphi_R$ | Correlation of recruitment deviations |
| $\sigma_{pe}$ | SD of process error |
| $L_\infty$ | Asymptotic length |
| $K$ | Von Bertalanffy growth rate |
| $L_0$ | Initial length |
| $cv_L$ | Coefficient of variation of length. |
| $\mu_{F_{a,y}}$ | Mean F |
| $\sigma_F$ | SD of F |
| $\varphi_{F_a}$ | Correlation of F between age classes |
| $\varphi_{F_y}$ | Correlation of F between years |
| *Random Effects* | |

| | | |
|---|---|---|
| $\delta_{R_y}$ | Recruitment deviation | |
| $\delta_{F_{a,y}}$ | Fishing mortality deviation | |
| $pe_{a,y}$ | Process error | |
| **Observation model** | | |
| $C_{ay}$ | Number of fish caught for each age class $a$ in year $y$ | |
| $P_{cla}$ | $P_{la}$ adding 0.5 to the ages to make it in the mid-year | |
| $E(I_{sly})$ | Expected survey index of abundance at length $l$ in a year $y$ | |
| $B_{ly}$ | Stock biomass for the length $l$ in a year $y$ | |
| $SW_l$ | Mean stock weight-at-length | |
| $CW_l$ | Mean catch weight-at-length | |
| $H_y$ | Harvest rate in year $y$ | |
| *Fixed Effects* | | |
| $\sigma_C$ | SD of the commercial catch | |
| $\varphi_C$ | AR(1) length correlation in catch | |
| $\sigma_I$ | SD of survey indices | |
| $\varphi_I$ | AR(1) length correlation in survey indices | |
| $\sigma_{q_{sl}}$ | SD of survey catchability | |
| *Random Effects* | | |
| $q_{sl}$ | Catchability of each survey | |

Table 3. ACL parameter estimates for the base model: couple Q's ≥45cm; Q40: couple Q's ≥ 40cm; Q30: couple Q's ≥30cm, and Q20: couple Q's ≥20cm. AIC is the Akaike information criterion, SD is the standard error.

| Model | Base | | Q40 | | Q30 | | Q20 | |
|---|---|---|---|---|---|---|---|---|
| AIC | 4911.93 | | 4892.23 | | 4878.20 | | 4910.00 | |
| | Estimate | SD | Estimate | SD | Estimate | SD | Estimate | SD |
| $\mu_{R_y}$ | 3267.6 | 1.468 | 3263.143 | 1.468 | 3126.103 | 1.467 | 2204.668 | 1.459 |
| $\sigma_R$ | 0.905 | 1.199 | 0.905 | 1.199 | 0.9 | 1.199 | 0.889 | 1.202 |
| $\sigma_{I_{fall}}$ | 0.724 | 1.045 | 0.724 | 1.045 | 0.73 | 1.045 | 0.761 | 1.045 |
| $\sigma_{I_{spring}}$ | 0.791 | 1.045 | 0.792 | 1.045 | 0.808 | 1.045 | 0.837 | 1.045 |
| $\sigma_{I_{spsh}}$ | 1.389 | 1.045 | 1.389 | 1.045 | 1.388 | 1.045 | 1.419 | 1.045 |
| $\varphi_I$ | 0.815 | 0.523 | 0.815 | 0.523 | 0.815 | 0.523 | 0.821 | 0.523 |
| $\sigma_C$ | 0.614 | 1.17 | 0.615 | 1.169 | 0.618 | 1.162 | 0.599 | 1.166 |
| $\varphi_C$ | 0.856 | 0.588 | 0.857 | 0.588 | 0.859 | 0.584 | 0.851 | 0.587 |
| $\sigma_{q_{fall}}$ | 0.463 | 1.212 | 0.463 | 1.212 | 0.452 | 1.212 | 0.379 | 1.218 |
| $\sigma_{q_{spring}}$ | 0.376 | 1.221 | 0.376 | 1.221 | 0.361 | 1.222 | 0.331 | 1.223 |
| $\sigma_{q_{spsh}}$ | 0.564 | 1.221 | 0.564 | 1.221 | 0.545 | 1.221 | 0.44 | 1.226 |
| $\sigma_{pe}$ | 1.027 | 1.103 | 1.024 | 1.103 | 1.083 | 1.1 | 1.035 | 1.101 |
| $K$ | 0.102 | 1.013 | 0.102 | 1.013 | 0.103 | 1.012 | 0.103 | 1.012 |
| $cv_L$ | 0.113 | 1.029 | 0.113 | 1.028 | 0.114 | 1.026 | 0.113 | 1.025 |
| $\sigma_F$ | 2.312 | 1.119 | 2.308 | 1.119 | 2.424 | 1.114 | 2.52 | 1.108 |
| $\mu_{F_{fish}}$ | 0.002 | 2.561 | 0.002 | 2.556 | 0.003 | 2.631 | 0.006 | 2.709 |
| $\mu_{F_{mort}}$ | 0.001 | 2.964 | 0.001 | 2.958 | 0.002 | 3.075 | 0.003 | 3.181 |

**Figures**

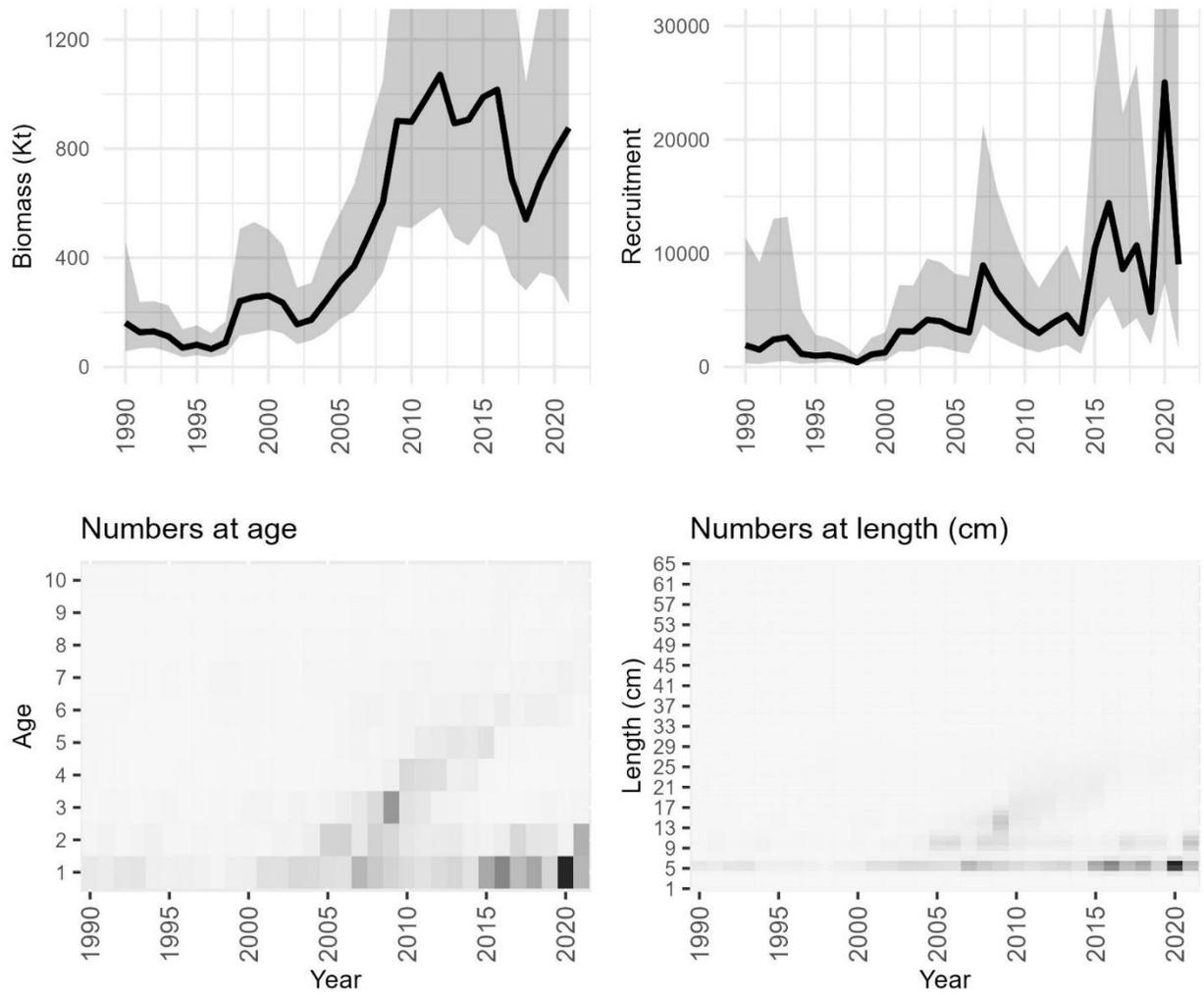

Figure 1. Population biomass (top left), recruitment (top right), numbers at age (bottom left) and numbers at length (bottom right). In the bottom panels, the darker the color, the larger the estimate.

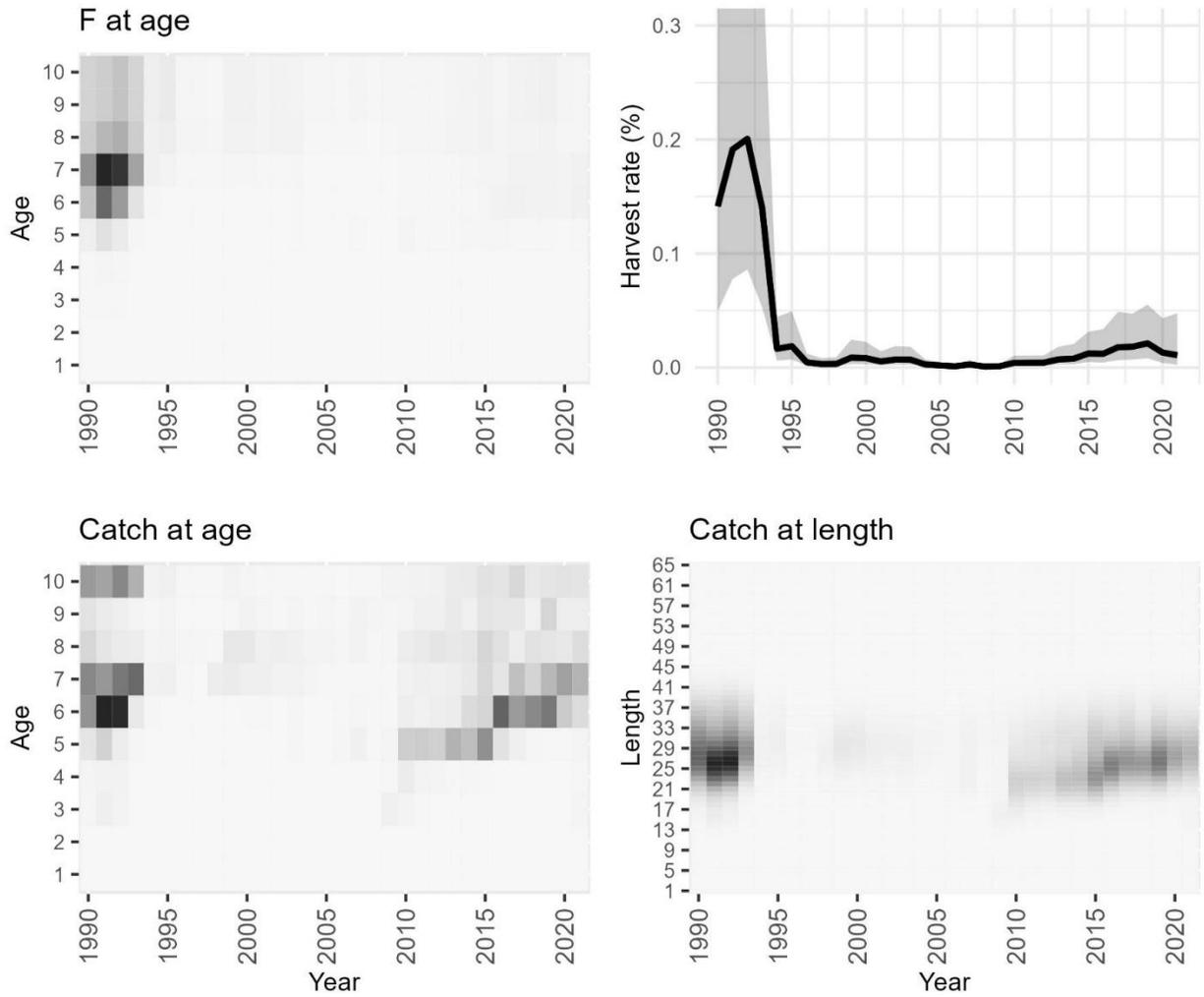

Figure 2. ACL model estimates of fishing mortality rates at age (top left), harvest rates (%, top right), catch at age (bottom left) and catch at length (bottom right) for 3LN redfish. In the top left and bottom panels, the darker the color, the larger the estimate.

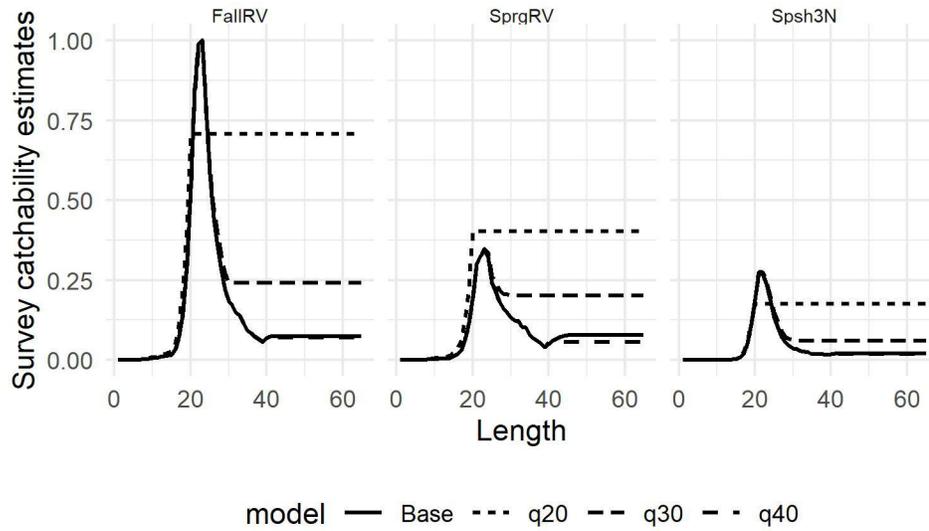

Figure 3. ACL model survey catchability estimates for the Canadian fall (FallRV), spring (SprgRV) and 3N EU-Spain (Spsh3N) surveys for various model formulations (base, q20, q30 and q40; see in text for details).

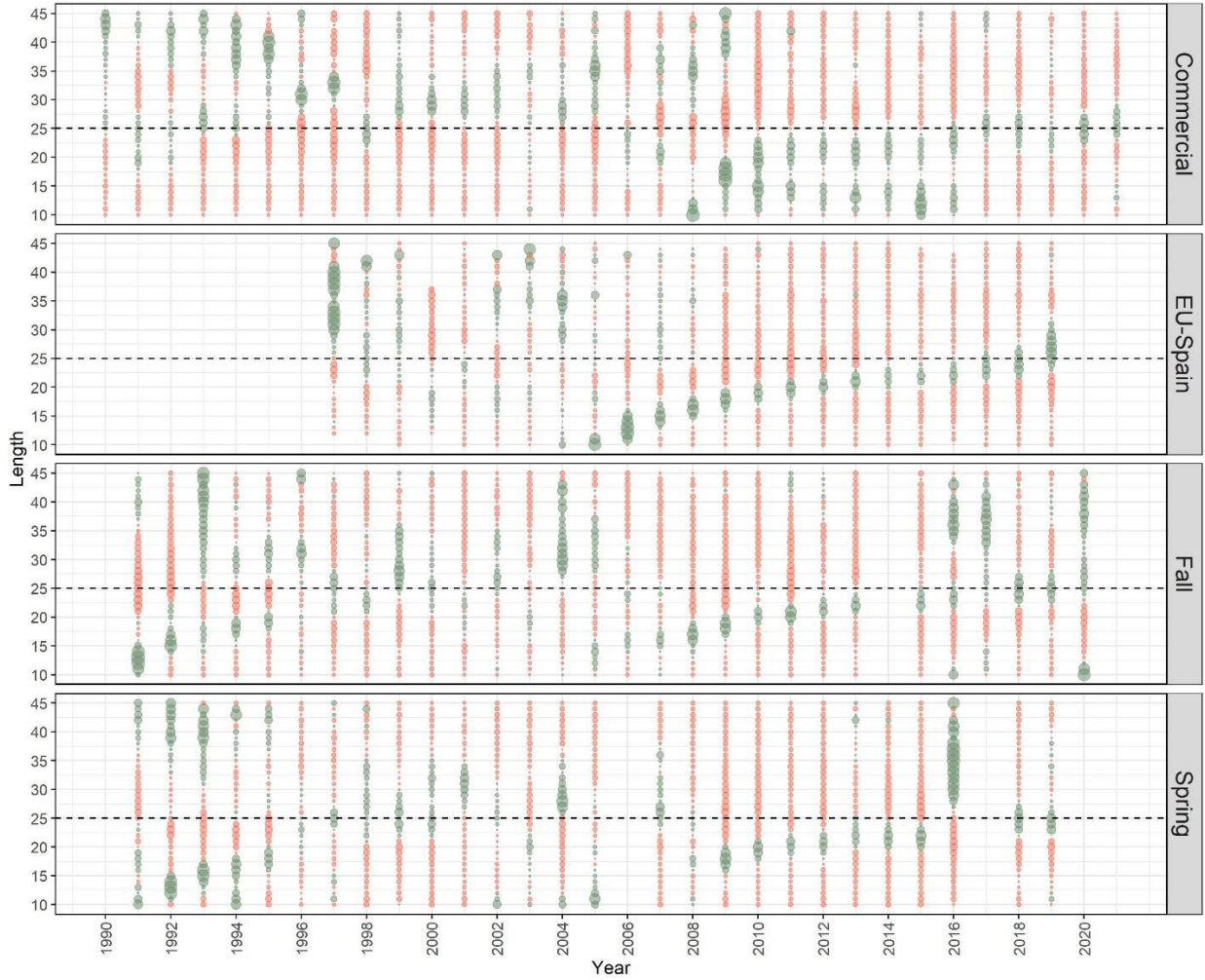

Figure 4. Standardized proportion at length and year plots (SPLY) for the commercial (top) and three survey length compositions. SPLY plots are useful for visualizing cohorts tracking through the catch. Unlike standard age-based (SPAY) plots, a cohort tracking through a SPLY plot will reach an asymptotic length, and trends in SPLY plots will more closely resemble a von Bertalanffy growth curve than a linear line.

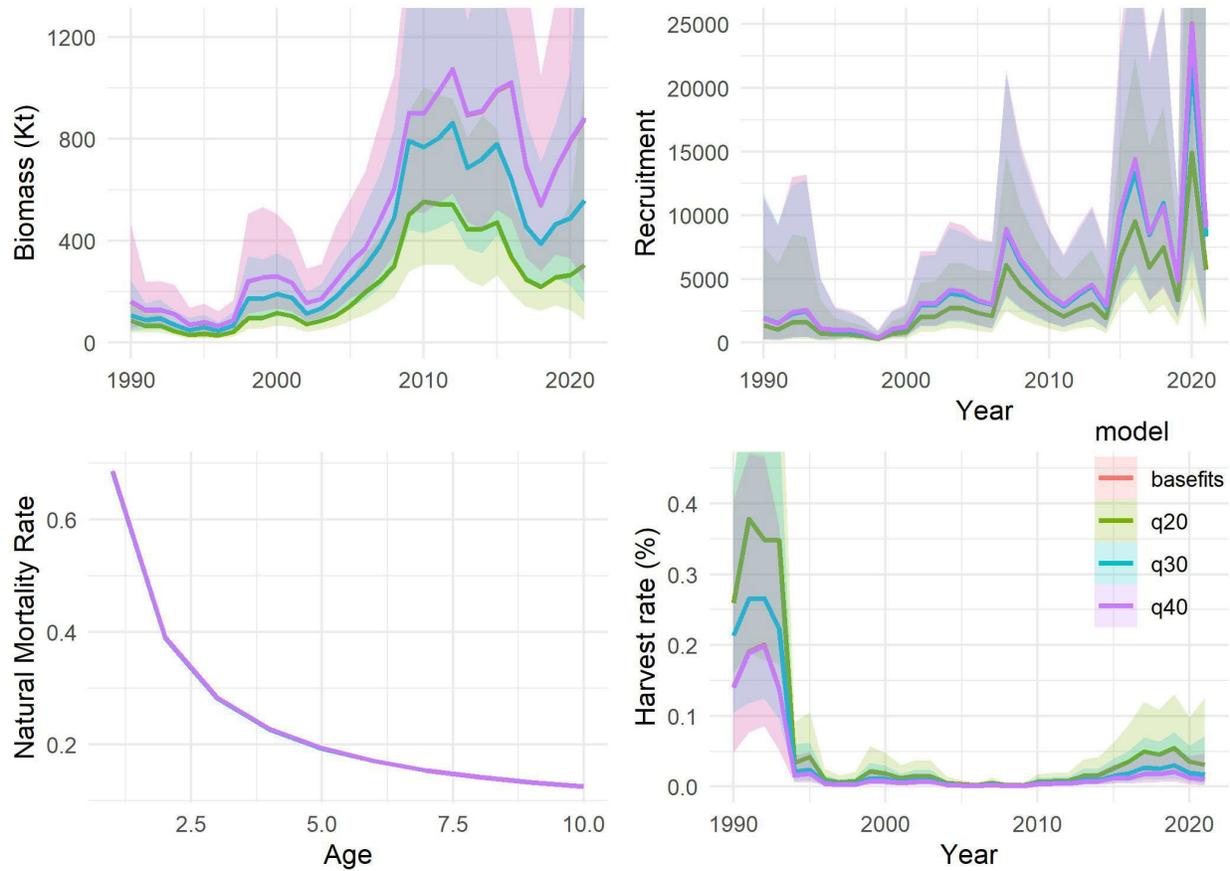

Figure 5. Comparison of population estimates from the survey catchability formulations (base case: fixing max estimate at 45cm; q35: fixing max estimate at 35cm; q25: fixing max estimate at 25cm, and q20: fixing max estimate at 20cm.

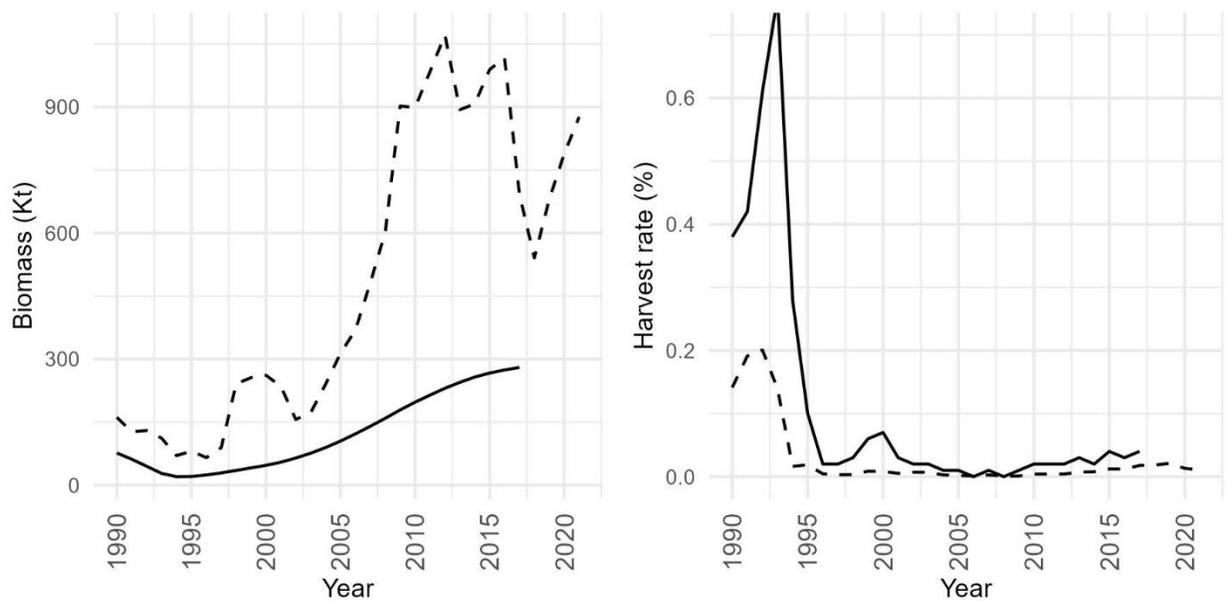

Figure 6. Comparison of biomass (left) and harvest rate (right) estimates from the ACL (dashed) and ASPIC (solid) stock assessment models for 3LN redfish.

**Appendix A**

**Supplementary Materials: A state-space catch-at-length assessment model for redfish on the Eastern Grand Bank of Newfoundland reveals large uncertainties in data and stock dynamics**

SF 1: Proportion of 3LN redfish survey area covered over time for the Canadian fall and spring and EU-Spain surveys in NAFO Divisions 3L and 3N.

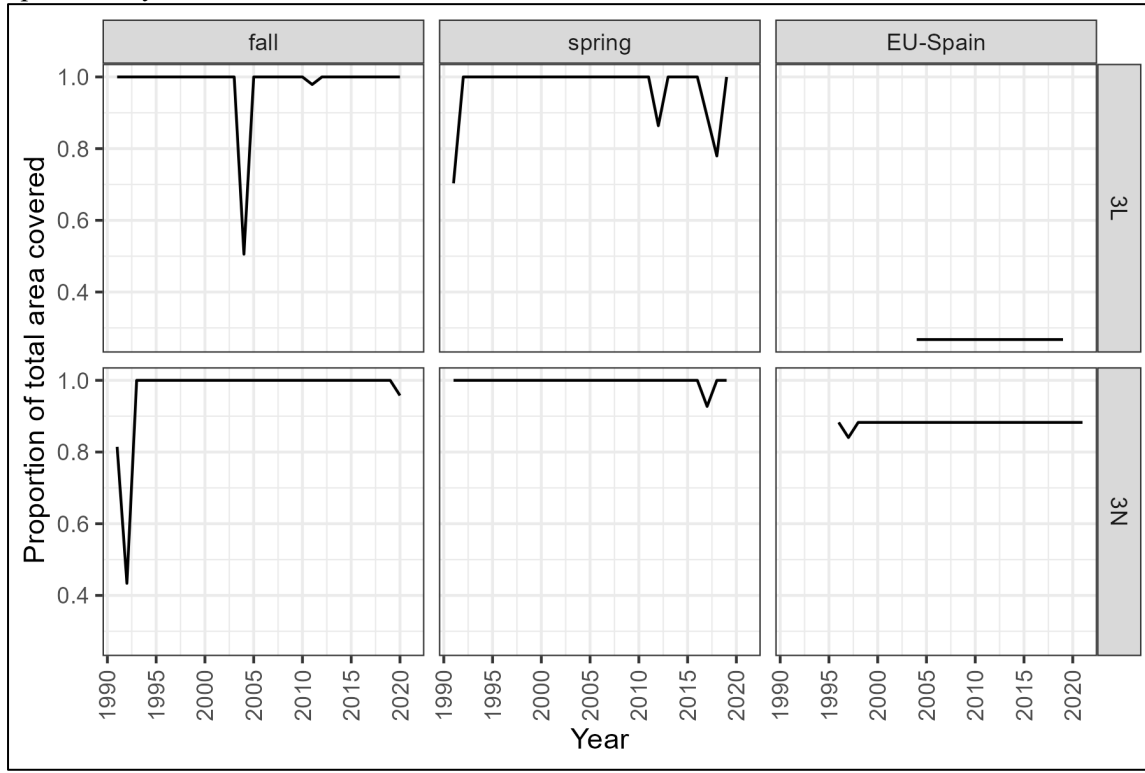

SF 2: Standardized residuals for the commercial catch at length fits for the base model run.

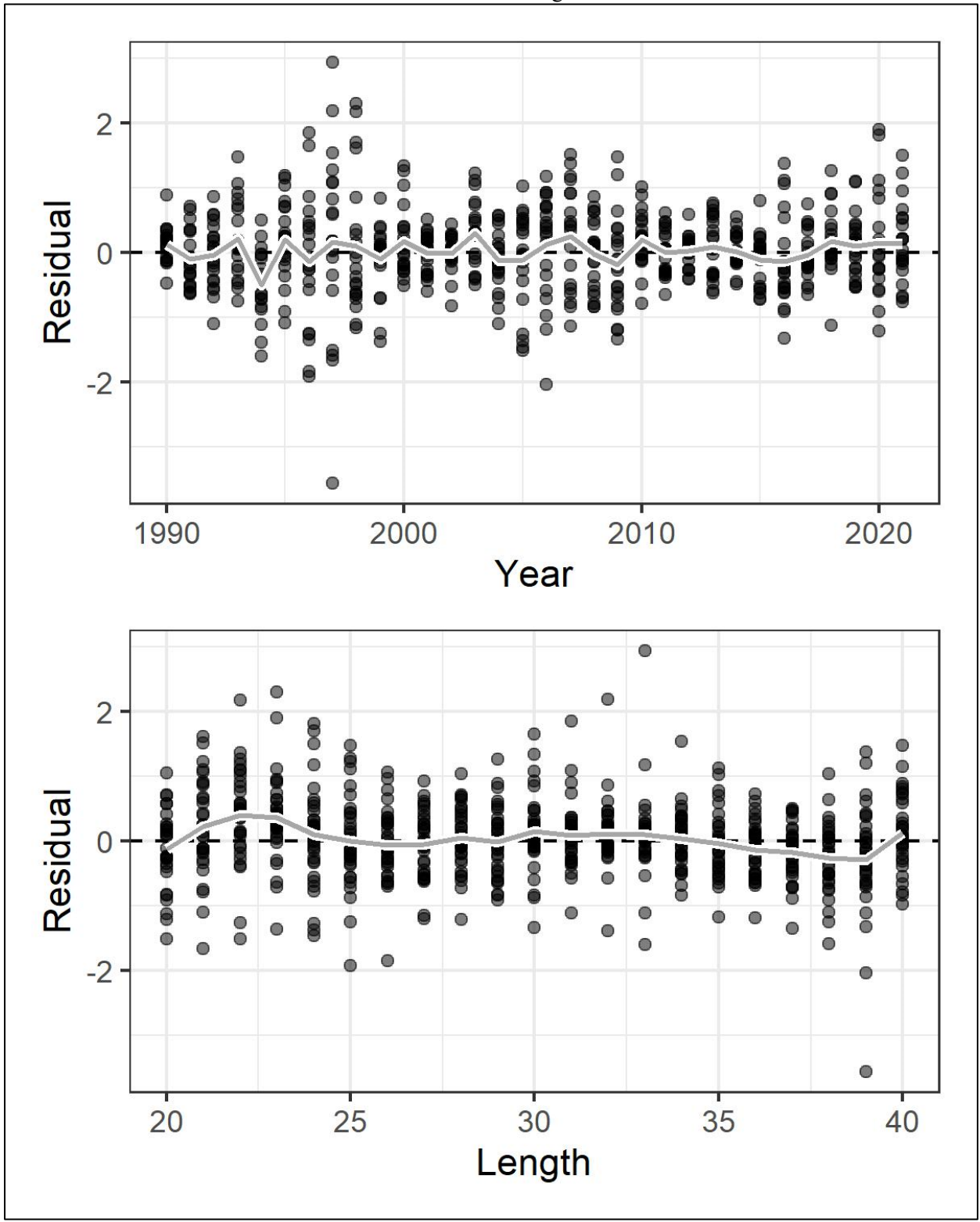

SF 3: Z residuals for the commercial catch at length fits for the base model run.

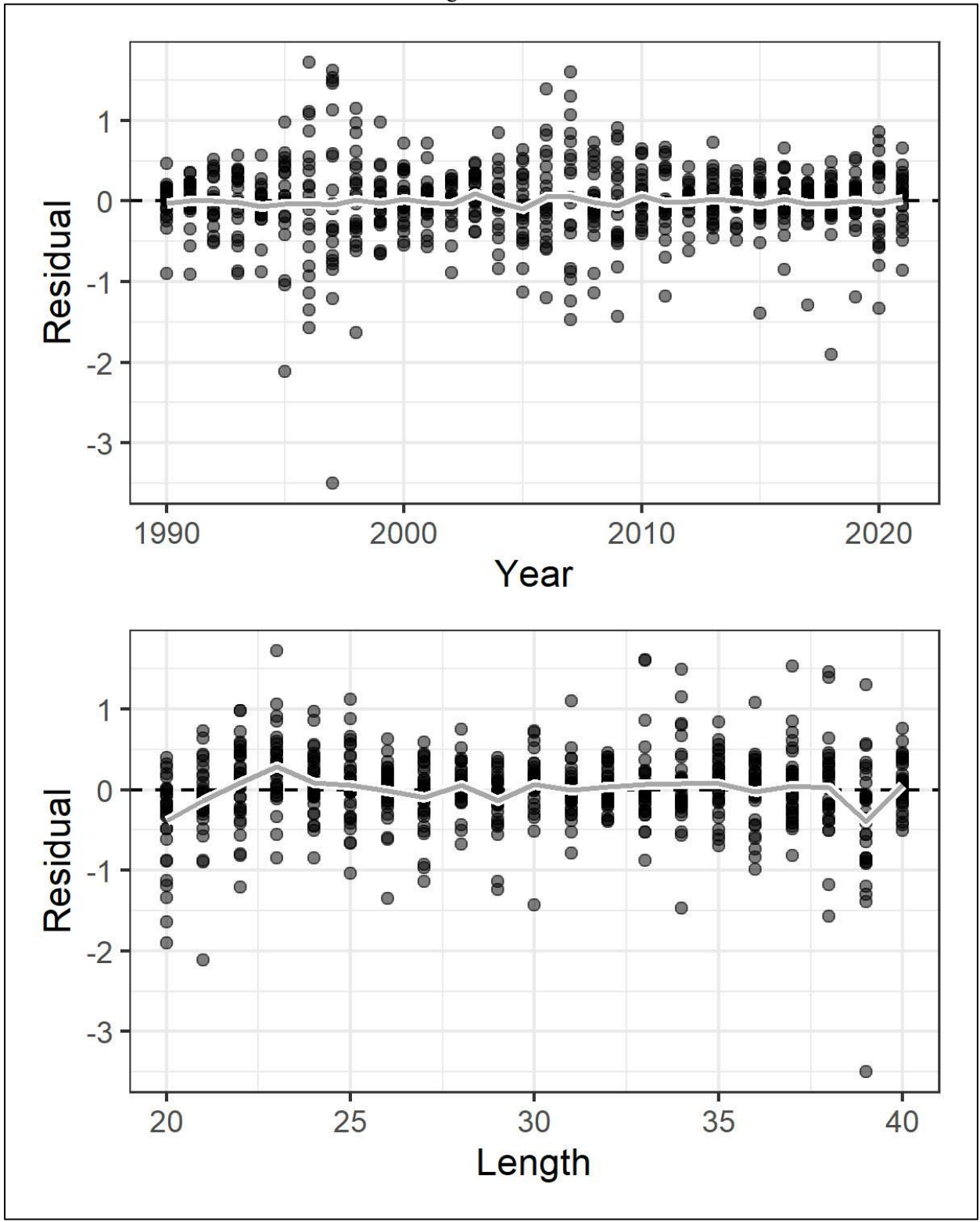

SF 4: Standardized residuals for the survey abundance at length fits for the base model run for the Canadian Fall (FallRV), Spring (SpringRV) and EU-Spain 3N (Spsh3N) surveys.

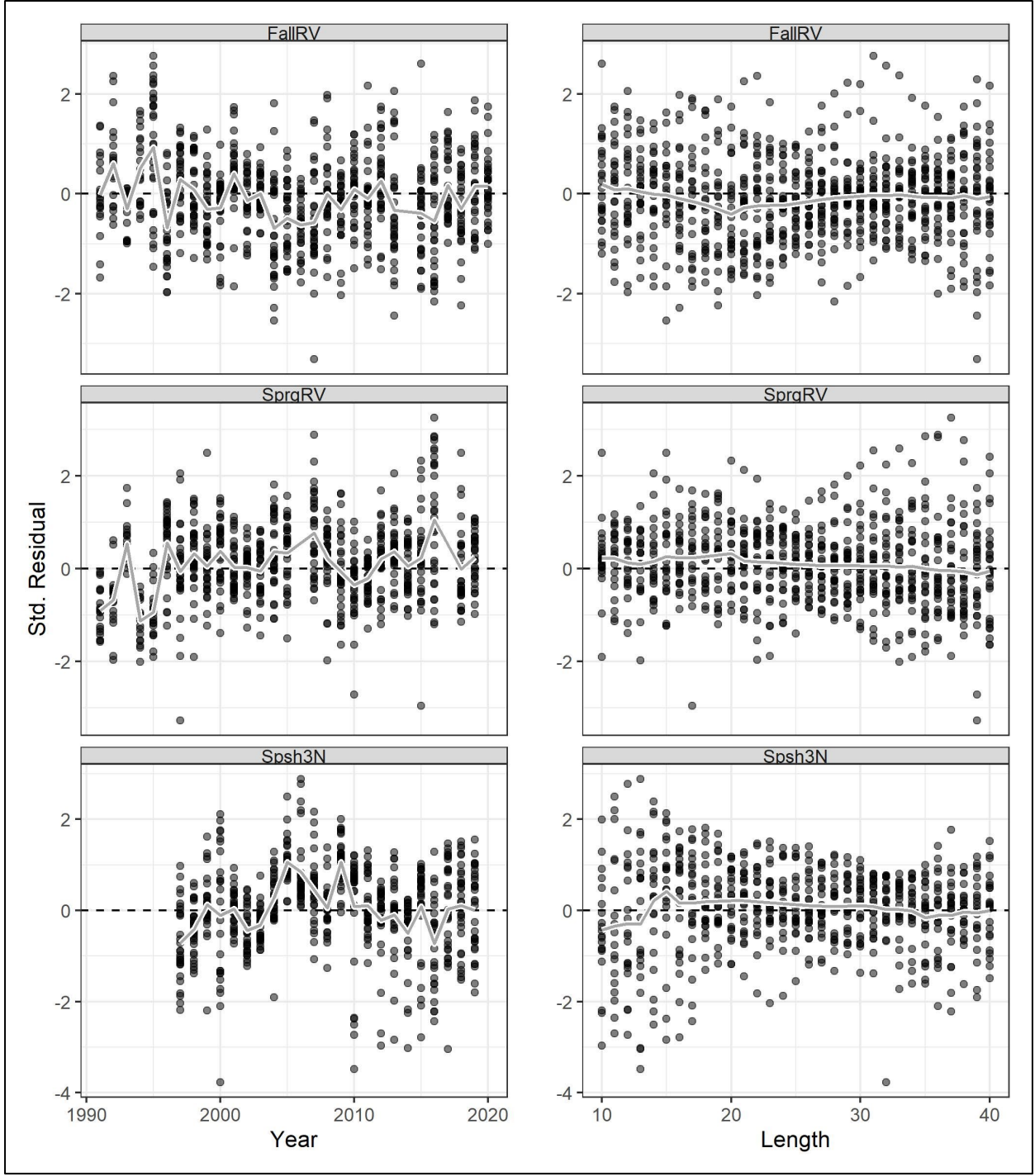

SF 5: Z residuals for the survey abundance at length fits for the base model run for the Canadian Fall (FallRV), Spring (SpringRV) and EU-Spain 3N (Spsh3N) surveys.

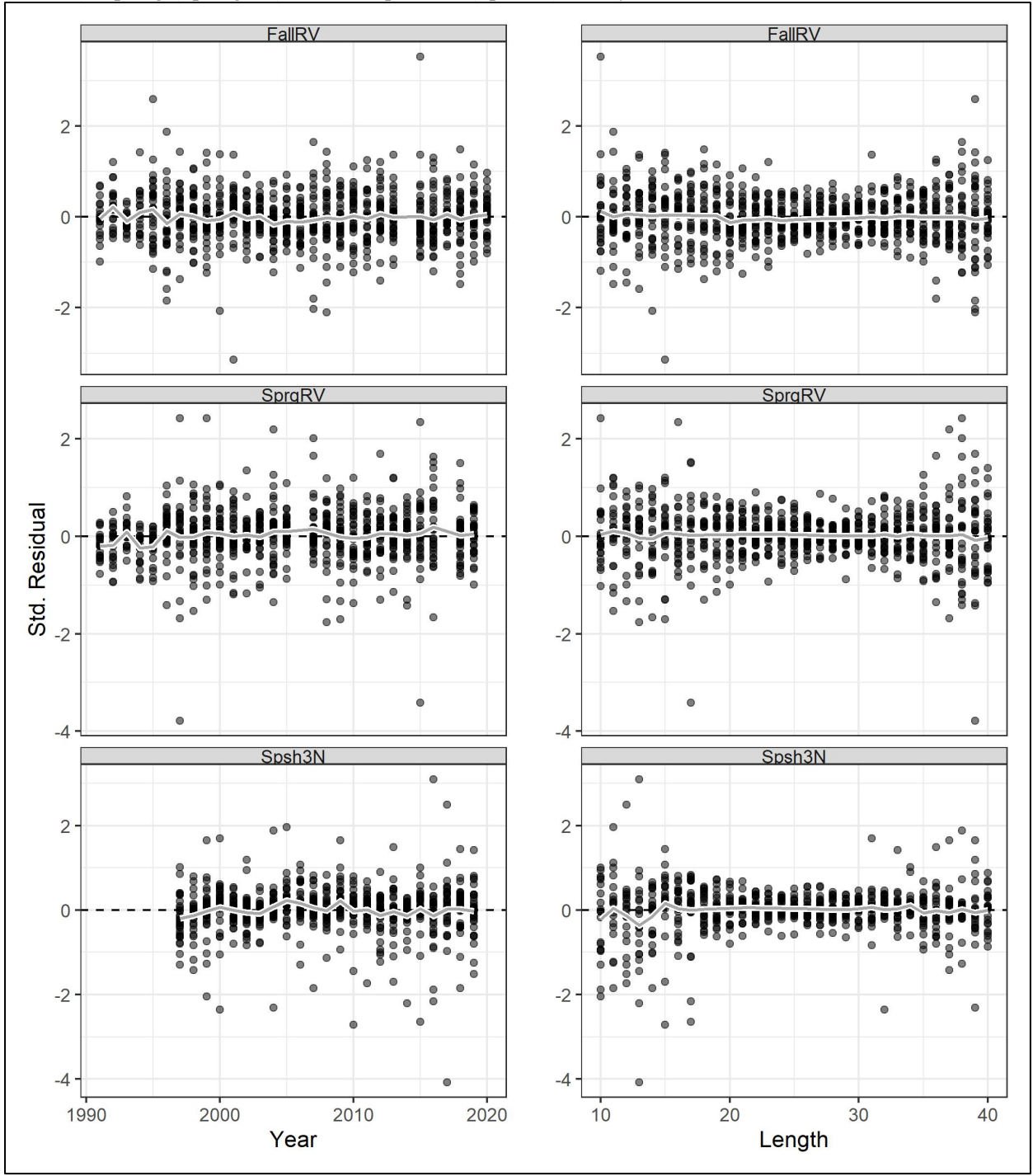

SF 6: Observed (left) and predicted (right) commercial catch at length from the ACL base run model.

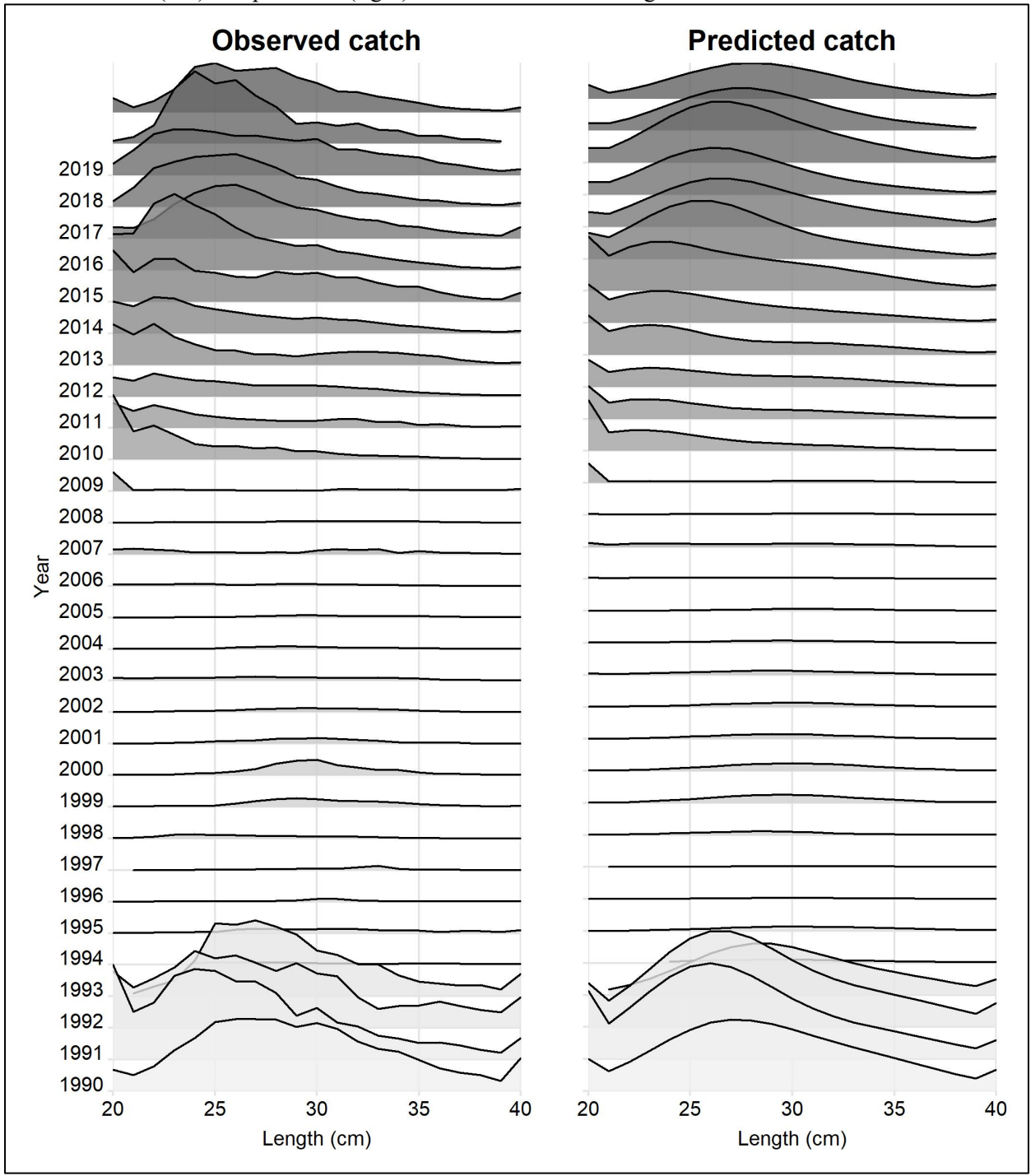

SF 7: Observed (left) and predicted (right) EU-Spain RV abundance at length from the ACL base run model.

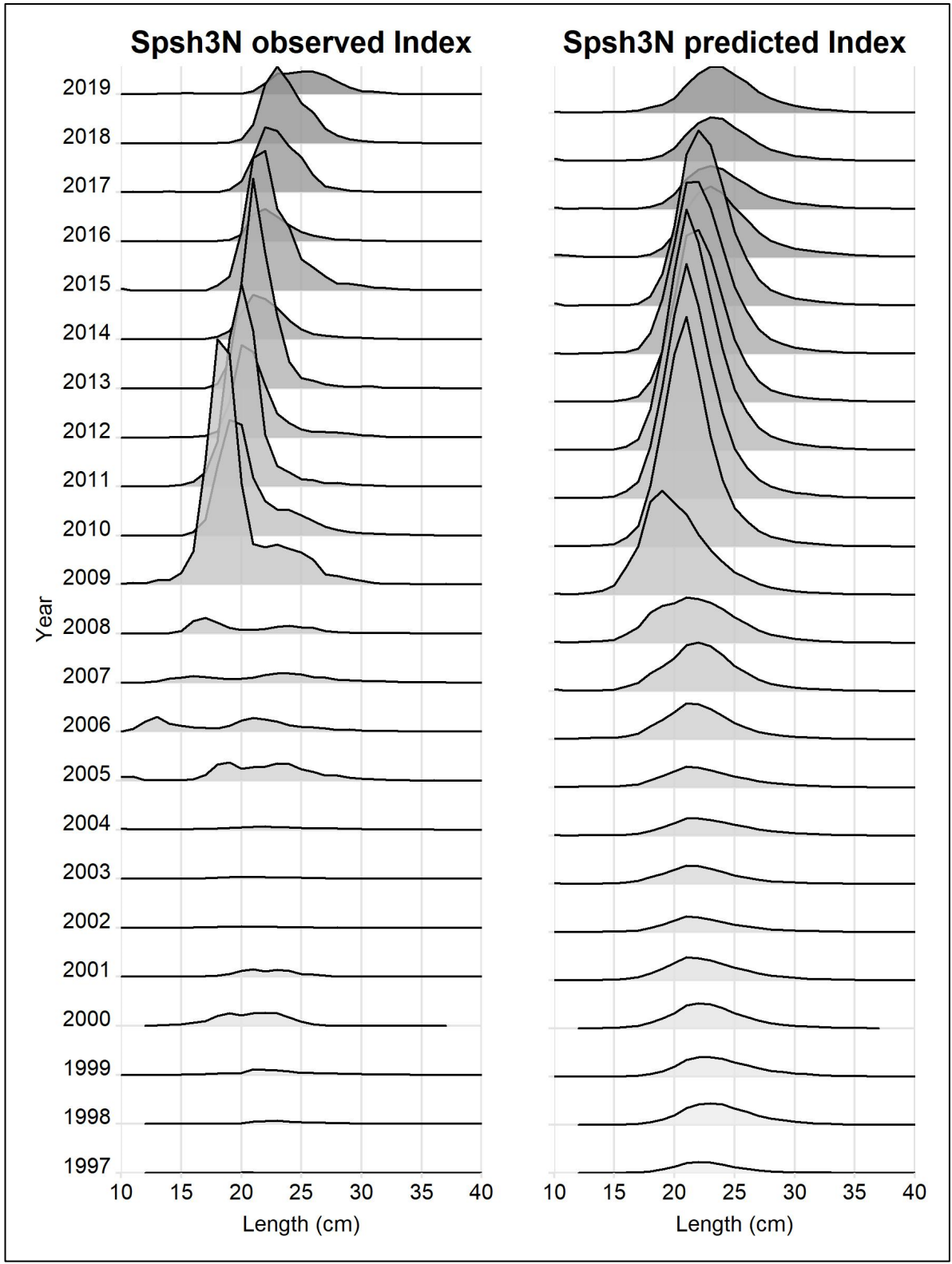

SF 8: Observed (left) and predicted (right) Canadian fall RV abundance at length from the ACL base run model.

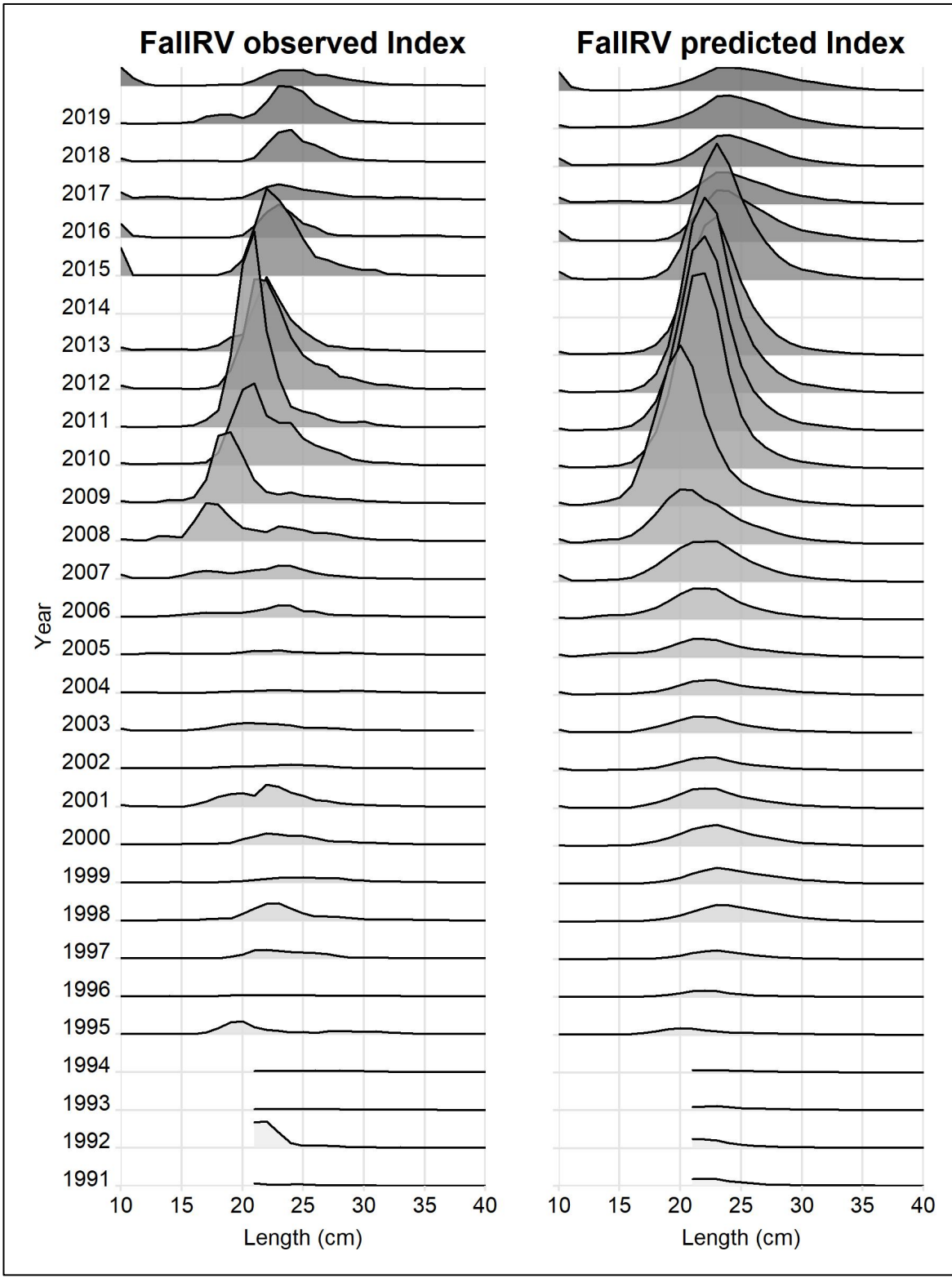

SF 9: Observed (left) and predicted (right) Canadian spring RV abundance at length from the ACL base run model.

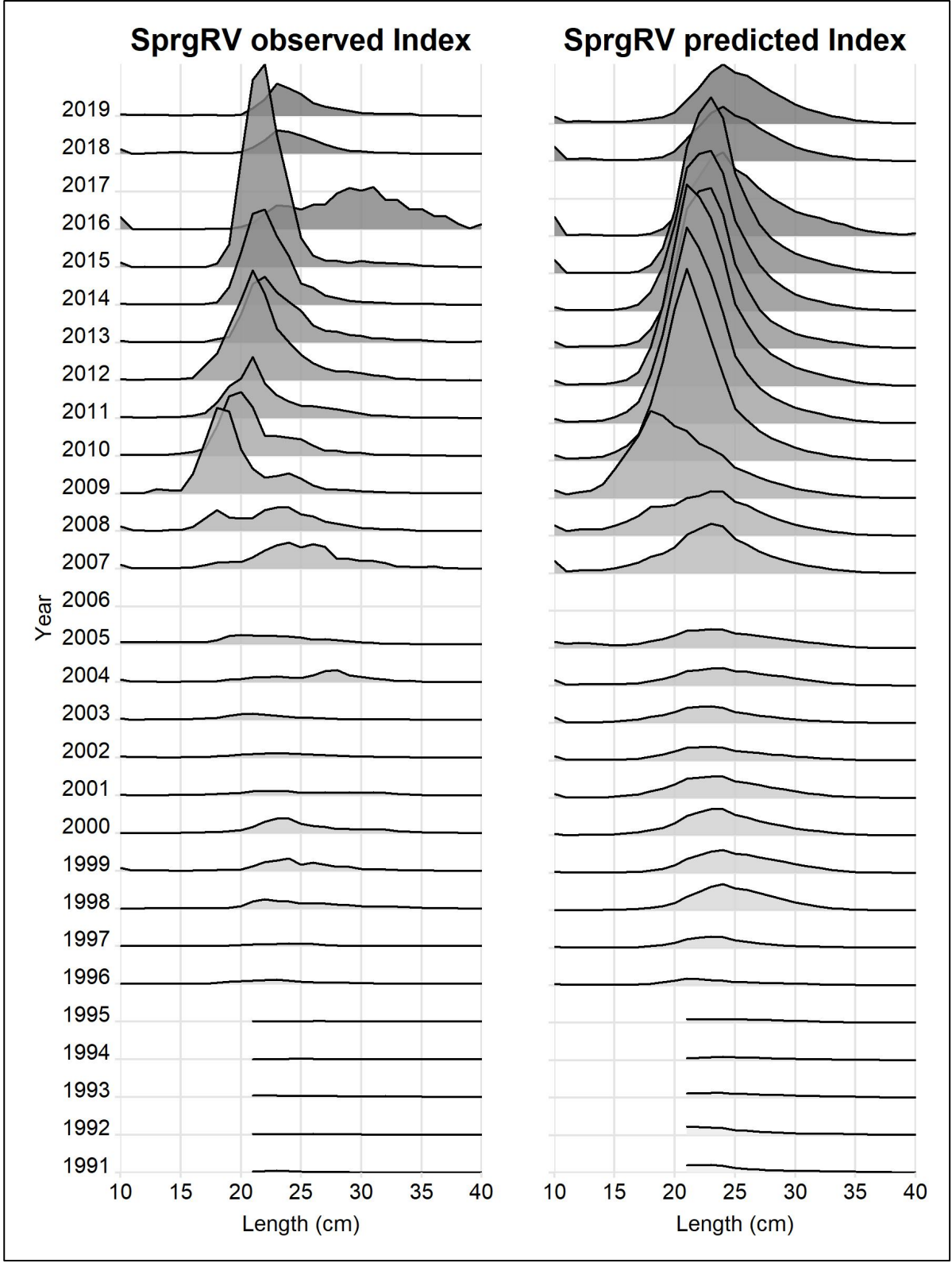

SF 10: Proportion of fish greater than 30cm in the total catch in each year by source (Commercial catch, green; Canadian fall RV, orange; Canadian spring RV, purple; EU-Spain RV, pink).

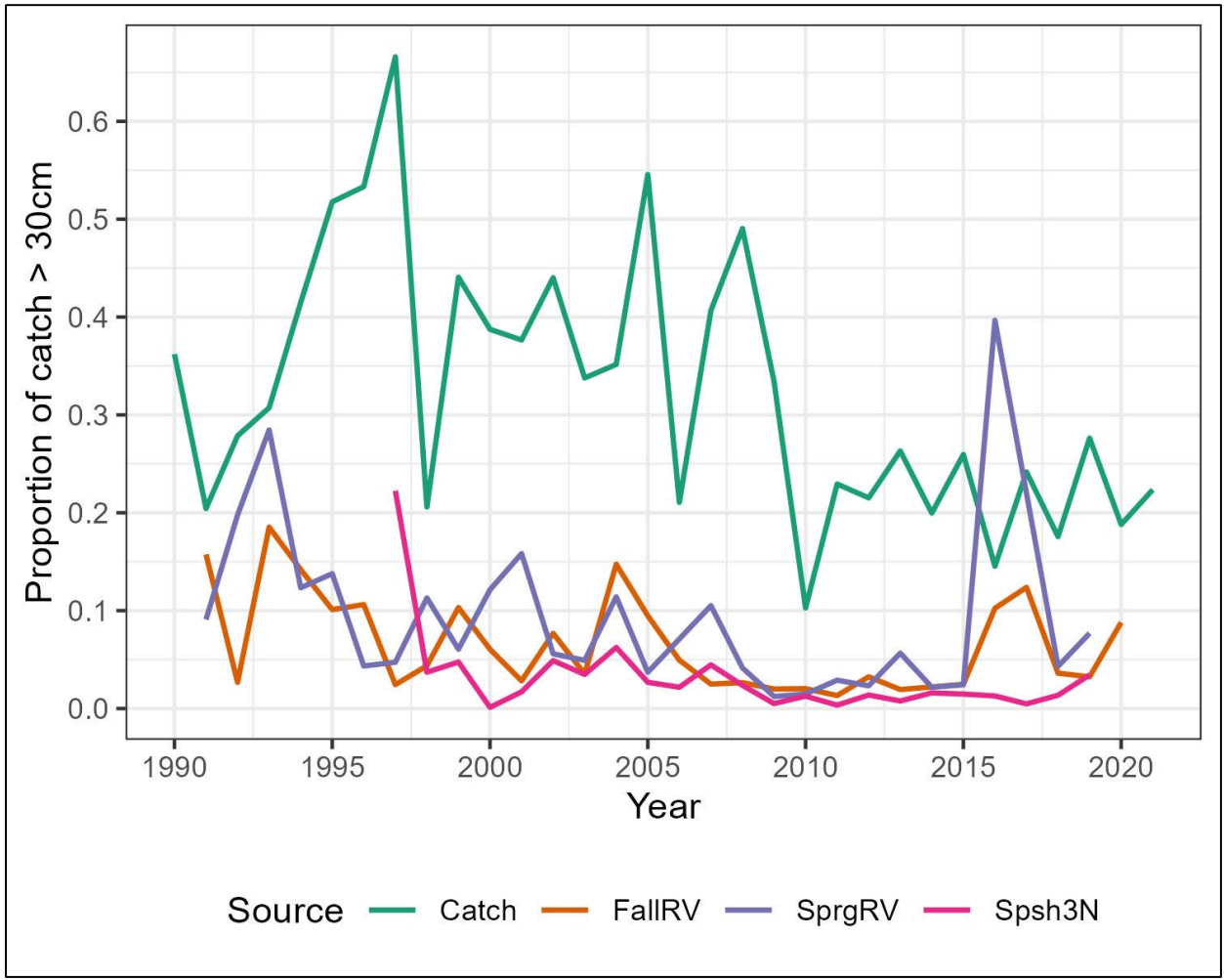

SF 11: ACL base model predicted process errors at year and age.

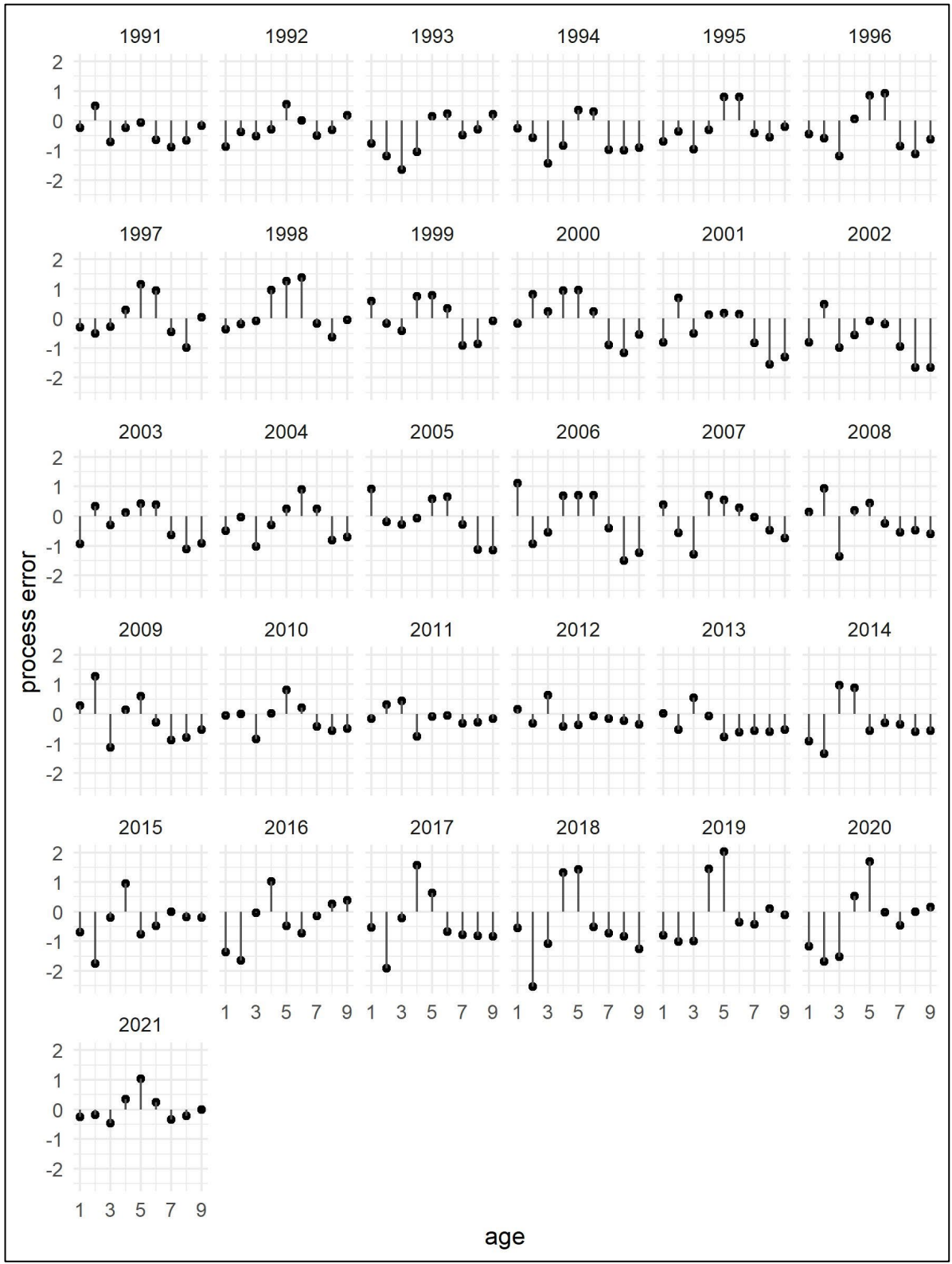

SF12: ACL base model bubble plots of predicted process errors. Red bubbles are positive and blue are negative.

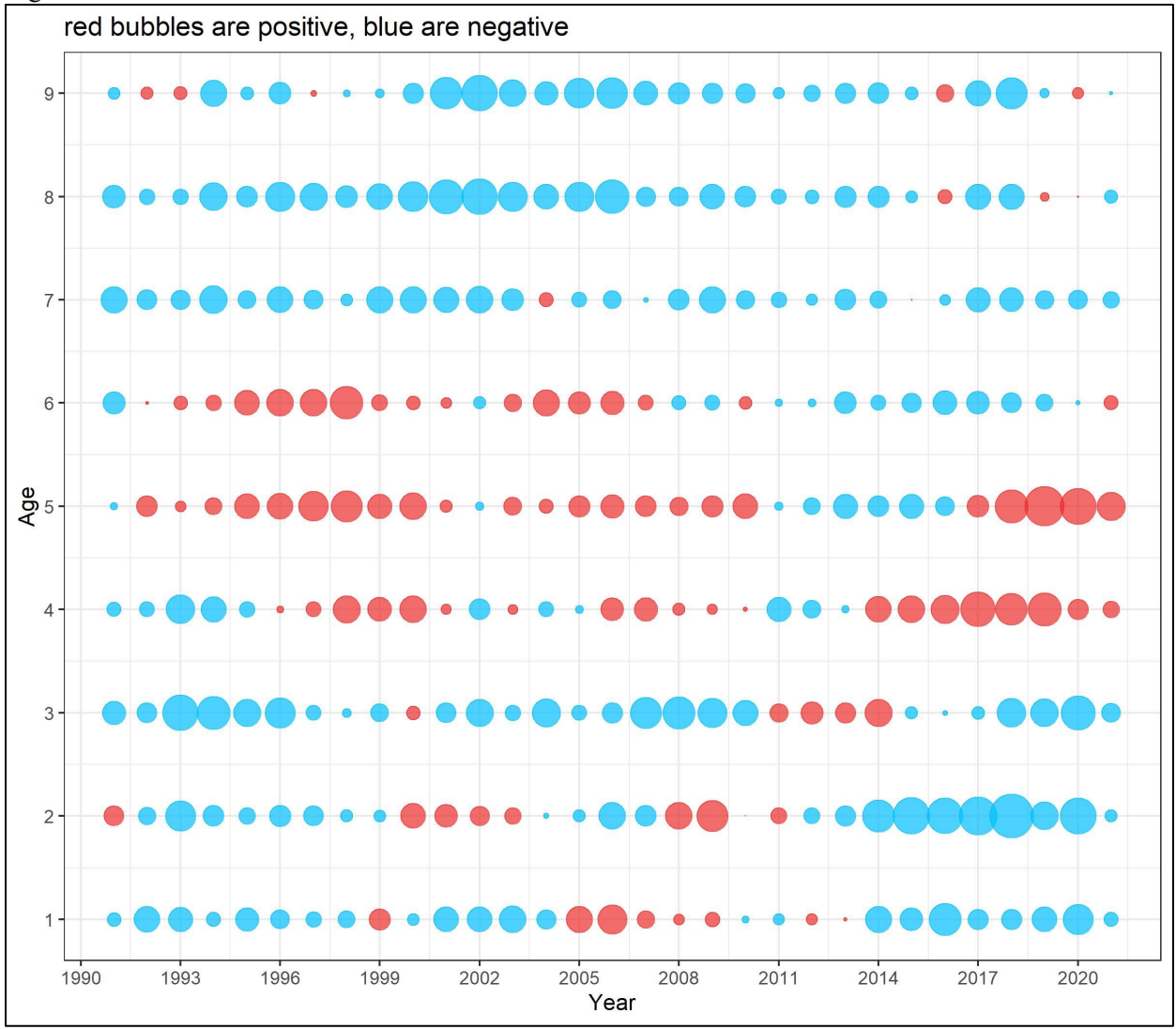